\documentclass[floatfix,prl,amsmath,twocolumn]{revtex4}
\usepackage{amssymb}
\usepackage{graphicx}
\usepackage{graphics}
\usepackage{epsfig}
\usepackage{psfrag}

\begin{document}

\newcommand{\be}{\begin{eqnarray}}
\newcommand{\ee}{\end{eqnarray}}
\newcommand{\bea}{\begin{eqnarray}}
\newcommand{\eea}{\end{eqnarray}}
\newcommand{\bma}{\begin{subequations}}
\newcommand{\ema}{\end{subequations}}
\def\qp{\pm}
\def\qed{\leavevmode\unskip\penalty9999 \hbox{}\nobreak\hfill
     \quad\hbox{\leavevmode  \hbox to.77778em{%
               \hfil\vrule   \vbox to.675em%
               {\hrule width.6em\vfil\hrule}\vrule\hfil}}
     \par\vskip3pt}
\def\lR{l^2_{\mathbb{R}}}
\def\RR{\mathbb{R}}
\def\E{\mathbf e}
\def\D{\boldsymbol \delta}
\def\S{{\cal S}}
\def\T{{\cal T}}
\def\dd{\delta}
\def\id{{\bf 1}}
\def\E{{\bf E}}
\newcommand{\Q}[1]{{{\footnotesize{\emph{[ #1 ] }}}}}
\newcommand{\M}[1]{{\marginpar{\footnotesize{\emph{ #1  }}}}}
\newcommand{\tr}[1]{{\rm tr}\left[#1\right]}
\newtheorem{theorem}{Theorem}
\newtheorem{prop}{Proposition}
\newtheorem{definition}{Definition}
\newtheorem{lemma}{Lemma}
\newcommand{\C}{{\Bbb C}}
\newtheorem{cor}{Corollary}
\title{Entanglement frustration for Gaussian states on symmetric graphs}

\author{
  M.M. Wolf, F. Verstraete, and J.I. Cirac}
\affiliation{\footnotesize  Max-Planck-Institut f\"ur
Quantenoptik, Hans-Kopfermann-Str. 1, Garching, D-85748, Germany}

\date{\today}

\begin{abstract}
We investigate the entanglement properties of multi-mode Gaussian
states, which have some symmetry with respect to the ordering of
the modes. We show how the symmetry constraints the entanglement
between two modes of the system. In particular, we determine the
maximal entanglement of formation that can be achieved in
symmetric graphs like chains, $2d$ and $3d$ lattices, mean field
models and the platonic solids. The maximal entanglement is always
attained for the ground state of a particular quadratic
Hamiltonian. The latter thus yields the maximal entanglement among
all quadratic Hamiltonians having the considered symmetry.
\end{abstract}

\maketitle

Classically as well as quantum mechanically the global ordering or
symmetry of a system often imposes highly non-trivial constraints
on its local properties. These kinds of frustration effects lie at
the heart of ordered interacting systems, and physicists are faced
with these phenomena whenever dealing with lattice systems or
molecular structures. The present paper is devoted to investigate
how the entanglement of two subsystems of a larger system is
constraint by such a global symmetry for some particularly
interesting class of states, the so--called Gaussian states
\cite{Hol82}.

Gaussian states appear very naturally in several branches of
physics where entanglement plays a predominant role. The
electromagnetic field in most quantum optical setups, atomic
ensembles interacting with such fields \cite{JKP00}, the motion of
a collection of trapped ions, or the low energy (bosonic)
excitations of many interacting systems can be very well described
by these states. This is due to the fact that quantum field
theories can be, in some regimes, approximated by Hamiltonians
which are quadratic in some bosonic operators, and thus in thermal
equilibrium as well as a result of the dynamics the corresponding
states are Gaussian. Thus, there is a growing interest in
understanding the entanglement properties of these states
\cite{normform,JKP00,WW01a,AEPW02,GWKWC03,WGKWC03}.

Our results quantify a very intuitive property of entanglement,
which distinguishes it from the usual correlations found in
classical systems: one particle can share entanglement only with a
limited number of other particles \cite{Wer89b}, which in turn
becomes smaller and smaller as the amount of entanglement
increases. Furthermore, the entanglement that can be shared by a
subset of particles strongly depends on the symmetries of the
multi-particle state. For example, if we have a set of particles
distributed on a lattice in a state with translational symmetry,
the maximal entanglement between any two nearest neighboring
particles should depend on the number of spatial dimensions, and
should decrease if the total number of particles increases. For
  one-dimensional rings of
 spin $\frac12$ particles quantitative investigations of
this kind were started in \cite{OW01}. However, the involved
optimization problems are highly non-trivial such that up to now
only a lower bound for the achievable {\it Entanglement of
Formation} (EoF) \cite{BDSW96} is known. In the case of Gaussian
states, the situation can become even more intriguing since for
two modes only, the amount of entanglement becomes unbounded. If
we consider three modes, and impose that the global state is
invariant under permutations, it turns out that the maximum EoF
between any pair of modes becomes finite. By increasing the number
of modes and imposing different symmetries to the global state,
this quantity experiences strong modifications. In this work we
determine the Gaussian state of $N$ modes which gives rise to the
maximal EoF between a selected pair of modes, for any $N$ and a
large variety of symmetry groups.

From our analysis it also follows that the state for which the
maximum entanglement is generated under a given symmetry
corresponds to the ground state of a particular Hamiltonian,
quadratic in the bosonic operators, which can be easily
constructed. Thus we can determine the Hamiltonian, invariant
under a certain symmetry group, that generates the maximum
two--mode entanglement for the physical systems mentioned above.

\begin{figure}[t]
\psfrag{E}{$E_G(\rho)$}
\psfrag{e}{\vspace*{6pt}$E_{max}$}\psfrag{n}{N}
\vspace*{-5pt}\epsfig{file=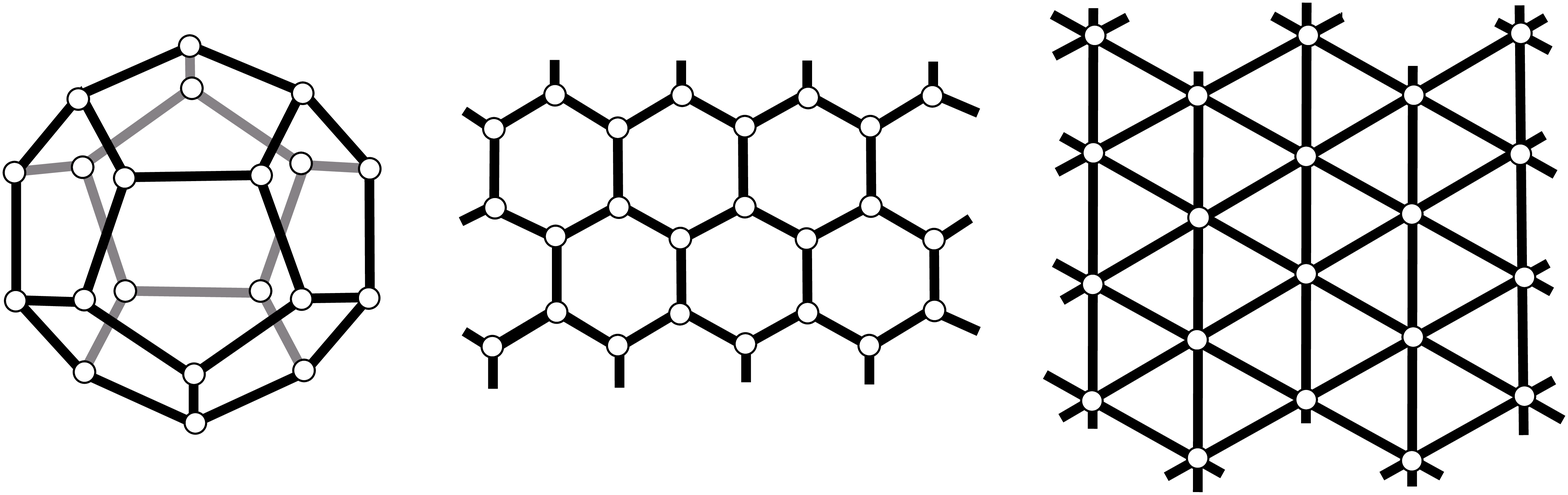,width=8.5cm}
\epsfig{file=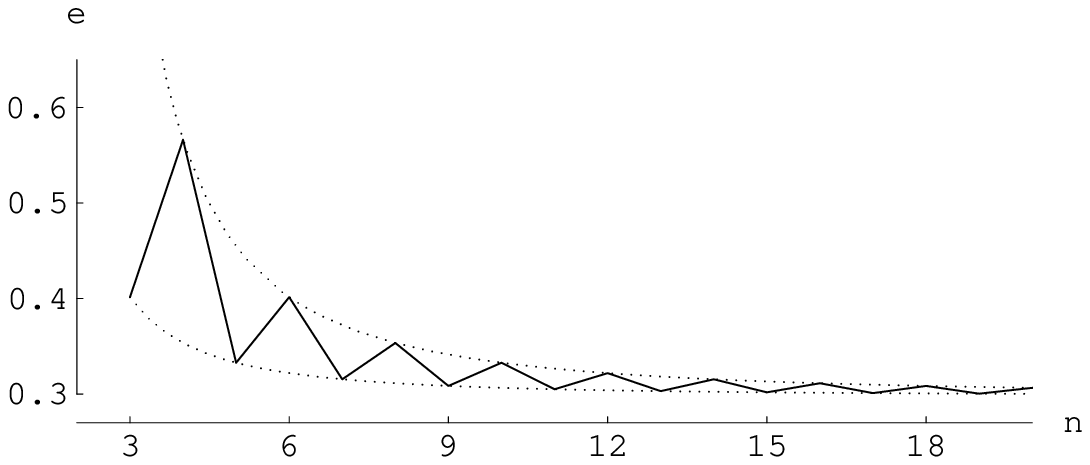,width=8.5cm} \caption{ (a): Apart from
chains, cubic lattices and meanfield clusters there are several
familiar symmetric graphs. Examples are the five platonic solids
(e.g. the dodecahedron), and hexagonal or trigonal lattices. (b):
Maximal nearest neighbor entanglement $E_{max}$ (ebits) in a ring
of $N$ harmonic oscillators. The dotted curves represent the
envelopes corresponding to Eq.(\ref{D0chain}).\label{doublefig}}
\label{fig:fiberzero}
\end{figure}

Although we will consider rather general symmetry groups, we will
illustrate our results for groups which can be associated to
symmetric graphs (Fig.\ref{doublefig}(a)) , since they give an
intuitive geometric depiction of the group and they are the ones
that naturally appear in many physical systems. For example, we
will give the optimal EoF for states that have the symmetries of a
lattice in any dimension, including square, cubic, hexagonal, and
trigonal lattices (Tab.\ref{tablattice}), or those of all platonic
solids (Tab.\ref{tabplato}).

Let $(Q_1,\ldots,Q_N,P_1,\ldots,P_N):=R$ be the $N$ conjugate
pairs of canonical operators characterizing $N$ modes and obeying
the canonical commutation relations $[R_k,R_l]=i\sigma_{kl}$ with
$\sigma$ being the symplectic matrix (cf.\cite{Hol82}).  Let us
consider a subgroup $G$ of the permutation group and two
particular modes, $k,l\le N$, for which there exists a group
element such that $g(k)=l$ and $g(l)=k$ (this condition will be
relaxed later on). We construct a Hamiltonian operator as follows
 \be
 \label{Hamiltonian}
 \hat{H}_{max}\hspace*{-4pt}&=&\hspace*{-2pt}\frac1{4|G|}\sum_{g\in G} \left(Q_{g(k)}+Q_{g(l)}\right)^2
 + \left(P_{g(k)}-P_{g(l)}\right)^2.
 \ee
Let us denote by $E_0$ the ground state energy and by $\Psi_0$ the
corresponding ground state, which is a Gaussian state, i.e.,
$\Psi_0$ has a Gaussian Wigner function. The relation between this
Hamiltonian and the EoF will later on be established by
linearizing the expression for the latter. We will show that the
Gaussian state which is invariant under $G$ and which maximizes
the EoF of the modes $k$ and $l$ is exactly $\Psi_0$ and that the
corresponding EoF is $E_{max}=E_F(E_0)$, where
 \begin{equation}\label{EoF}
  E_F(\Delta)=c_+(\Delta)\log[c_+(\Delta)]-
 c_-(\Delta)\log[c_-(\Delta)],
 \end{equation}
and $c_{\pm}(\Delta):= ( \Delta^{-1/2}\pm \Delta^{1/2})^2/4$.
Hence, the ground state of $\hat{H}_{max}$ has maximum
entanglement under all $G$-invariant Gaussian states and
conversely, among all quadratic $G$-invariant Hamiltonians
$\hat{H}_{max}$ generates the largest amount of entanglement at
zero temperature.

The ground state of Hamiltonians of the form (\ref{Hamiltonian})
can be easily determined by resorting to symplectic space. We
define the covariance matrix (CM) of a Gaussian state $\rho$ as
usual,
 \begin{equation}\label{Gamma}
 \Gamma_{kl}:=\Big\langle \{\big(R_k-\langle
 R_k\rangle\big),\big(R_k-\langle
 R_k\rangle\big)\}_+\Big\rangle,
 \end{equation}
which must fulfill $\Gamma\geq i\sigma$ \cite{Hol82}. Let us also
introduce the {\it Hamiltonian matrix} \cite{AEPW02} corresponding
to $\hat{H}_{max}$ as $H=H_+\oplus H_-$ with \footnote{We use
$|k\rangle$ to denote the unit vector in $\RR^N$ whose only
nonzero component is the $k$--th. Thus,
$T_g|k\rangle=|g(k)\rangle$. }
 \begin{eqnarray}\label{Hqptwirl}
 H_{\qp}&:=&\frac1{|G|}\sum_{g\in G} T_g\; h_{\pm}^{(k,l)}\;
 T_g^{-1},\\
 h_{\pm}^{(k,l)} &:=& \frac14\Big[|k\rangle\langle k
 |+|l\rangle\langle l |\pm(|k\rangle\langle l |+|l\rangle\langle k
 |)\Big].
 \end{eqnarray}
The matrix $H\geq 0$ can be diagonalized by a symplectic matrix
$S$, $H=SDS^T$. Since $\tr{\rho\hat{H}}=\tr{\Gamma H}$, the ground
state energy of $\hat{H}$ is given by
\footnote{Eq.(\ref{gsenergy}) is derived by exploiting that the
entries of the diagonal matrix $D$ are the square roots of the
eigenvalues of $(\sigma H \sigma^T H)$ \cite{Hol82}.}
 \begin{equation}\label{gsenergy}
 E_0= \inf_{\Gamma} \tr{\Gamma H} =\inf_{\Gamma} \tr{\Gamma D} =
 2||\sqrt{H_+^{1/2} H_- H_+^{1/2}}||_1\!.
 \end{equation}
Since $H$ has a null space, the CM $\Gamma_0$ of the ground state
$\Psi_0$ of $\hat H$ is achieved  in the limit $\epsilon\to 0$ of
$\Gamma_\epsilon:=S_\epsilon^{-1T}S_\epsilon^{-1}$, where
$S_\epsilon$ is the symplectic matrix that diagonalizes
$H+\epsilon\id$.

Before we prove the above statements, let us utilize the results
to analyze the maximum EoF for several interesting symmetry
groups. We will concentrate on groups which can be associated to a
symmetric graph.

Consider a simple undirected graph with $N$ vertices,
characterized by an {\it adjacency matrix} $A$, which is such that
$A_{kl}=1$ if the vertices $k$ and $l$ are connected by an edge,
and $A_{kl}=0$ if there is either no edge or $k=l$. The symmetry
group $G$ of the graph contains all permutations $g$, which
commute with the adjacency matrix $[A,g]=0$. The graph is called
{\it symmetric} if all edges as well as all vertices are equal in
the sense that every edge and every vertex can be mapped onto
every other one by an element of $G$. Examples of symmetric graphs
are given in Fig.\ref{doublefig}(a) and
Tab.\ref{tablattice},\ref{tabplato}. By utilizing this symmetry we
can simplify
 \begin{equation}\label{Hqptwirl2}
 H_{\qp} =
 \frac1{|{\cal E}|}\sum_{(k,l)\in{\cal E}} h_{\qp}^{(k,l)},
 \end{equation}
where ${\cal E}=\{(k,l)|A_{kl}=1\}$ is the set of edges. Hence,
the sum in Eq.(\ref{Hamiltonian}) runs over all edges which
correspond therefore to physical interactions between the adjacent
modes. By observing that $[H_+,H_-]=0$ we can derive the ground
state CM (or the respective limit):
 \begin{equation}\label{gamma0}
 \Gamma_0=\sqrt{H_-H_+^{-1}}\oplus \sqrt{H_+H_-^{-1}}.
 \end{equation}
Note that when acting on two modes only, the ground state of
$\hat{H}_{max}$ is the original singular EPR-state. Moreover,
since $H_-$ has a Kernel containing the vector $(1,1,\ldots,1)$
the ground state of any $\hat{H}_{max}$ will always be singular
and the maximal entanglement is thus only attained exactly in the
limit of infinite squeezing.

\begin{table}
\begin{tabular}{|l|c|c|}
  \hline
  lattice & $E_{max}$ & $N_a$ \\
  \hline
  hexagonal (2d) & 10.61 & 3 \\
  square (2d)& 6.31 & 4 \\
  trigonal (2d) & 2.69 & 6 \\
  cubic (3d) & 2.62 & 6 \\
  \hline
\end{tabular}
\caption{\label{tablattice}Maximal nearest neighbor entanglement
$E_{max}$ (in units of $10^{-2}$ ebits) for some infinite $2d$ and
$3d$ lattices. $N_a$ is the number of adjacent
vertices.\vspace*{-10pt}}
\end{table}

In the following we will apply the obtained results to some
examples of familiar symmetric graphs.

\vspace*{-12pt}\subparagraph{Chains and rings:} The simplest
non-trivial example of a symmetric graph is a ring of $N$ nodes
representing translation and reflection symmetry. In this case the
operators $H_{\qp}$ have the form
\begin{equation}\label{Hqpchain2}
H_{\qp}=\frac1{4N}\big[2\id\pm(T+T^{-1})\big]
\end{equation}
where $T_{kl}=\delta_{k,l+1 \mbox{mod} N}$ is the cyclic shift
operator. $H_{\qp}$ are so-called {\it circulant matrices}
\cite{HJ87}, which can be  diagonalized simultaneously by a
Fourier transform yielding
 \begin{equation}\label{D0chain}
 E_0=\frac1N\sum_{l=0}^{N-1} \Big|\sin{\frac{2\pi}N l}\Big|
=\left\{\begin{array}{cc}
 \frac2N\cot{\frac{\pi}N}, &N\ \mbox{even}  \\
 \frac1N\cot{\frac{\pi}{2N}}, &N\ \mbox{odd}
 \end{array}\right..
 \end{equation}
Hence, the entanglement  remains finite and is suppressed in rings
with an odd number of modes (see Fig.\ref{doublefig}). It
approaches $E_{max}=0.30$ ebits ($E_0=\frac2{\pi}$) in the limit
$N\rightarrow\infty$. This value is comparable to the $0.29$ ebits
conjectured for an infinite chain of spin $\frac12$ particles
\cite{OW01}.

\vspace*{-12pt}\subparagraph{Cubic lattices:}The result obtained
for the chain has a straight forward extension to $d$-dimensional
cubic lattices. Imposing periodic boundary conditions (i.e. a
lattice on a torus) for a cubic lattice of $N^d$ modes we get
 \begin{equation}\label{Hqpchaind}
 H_{\qp}=\frac1{4N^d}\Big[2\id\pm\frac1d\sum_{a=1}^d(T_{(a)}+T_{(a)}^{-1})\Big],
 \end{equation}
where now $T_{(a)}$ is the shift operator acting on the $a$'th of
$d$ tensor factors, each corresponding to one of the dimensions of
the lattice (e.g. $T_{(2)}=\id\otimes T\otimes\id\ldots$).
Diagonalizing $H_{\qp}$ by a tensor product of Fourier transforms
leads to
 \begin{equation}\label{D0cube}
 E_0=\frac1{N^d}\sum_{l_1=1}^N\ldots\sum_{l_d=1}^N
 \left[1-\frac1{d^2}\Big(\sum_{k=1}^d\cos\frac{2\pi}N
 l_k \Big)^2\right]^{\frac12},
 \end{equation}
which goes to 1 ($E_{max}\rightarrow 0$) for $d\rightarrow\infty$,
and is calculated explicitly in Tab.\ref{tablattice} for the
infinite two and three dimensional cubic lattice.

\vspace*{-12pt}\subparagraph{Mean field clusters:}When every mode
is connected to every other one, i.e., when we have complete
permutation symmetry, then
 \begin{equation}\label{Hqpperm}
 H_{\qp}=[2N(N-1)]^{-1}\big[(N-1)\id\pm({\bf E}-\id)\big],
 \end{equation}
where ${\bf E}_{kl}=1$, which leads to $E_0=\sqrt{\frac{N-2}N}$.
Hence, the maximal entanglement decreases with the number $N$ of
modes  and vanishes as $\sim \frac1{N^2}\log{N}$ in the limit
$N\rightarrow\infty$.

\begin{table}
\begin{tabular}{|l|c|c|c|c|}
  \hline
  platonic solid & $E_{max}$ & $N_a$ & $N$ & $E_0$ \\
  \hline
  tetrahedron & 19.74 & 3 & 4 & $\frac1{\sqrt{2}}$ \\
  cube & 19.74 & 3 & 8 & $\frac1{\sqrt{2}}$ \\
  dodecahedron & 11.12 & 3 & 20 & $\frac1{30}\big(12+5\sqrt{2}+2\sqrt{5}\big)$ \\
  octahedron & 10.75 & 4 & 6 & $\frac16(3+\sqrt{3})$ \\
  icosahedron & 5.37 & 5 & 12 & $\frac1{\sqrt{5}}+\frac1{\sqrt{6}}$ \\
  \hline
\end{tabular}
\caption{\label{tabplato}Maximal amount of nearest neighbor
entanglement $E_{max}$ (measured in units of $10^{-2}$ ebits) and
the respective ground state energy (minimal EPR-uncertainty) $E_0$
for the five platonic solids. $N_a$ is the number of adjacent
vertices and $N$ the total number of nodes.\vspace*{-10pt}}
\end{table}

\vspace*{-12pt}\subparagraph{Platonic solids:} The results for the
graphs corresponding to the three dimensional platonic solids can
be found in Tab.(\ref{tabplato}).

All these examples indicate three different tendencies for the
maximal EoF:\vspace*{-2pt}
\begin{enumerate}
    \item $E_{max}$ decreases with the number of adjacent vertices.\vspace*{-5pt}
    \item $E_{max}$ decreases with the total number of  vertices.\vspace*{-5pt}
    \item $E_{max}$ is suppressed in loops with an odd number of
    vertices, which give rise to  additional frustration. \vspace*{-3pt}
\end{enumerate}

Let us now to proceed to prove our main result. We denote by
$\Gamma$ and $\gamma$ CM of the global state and the reduced
density operator for the modes $k,l$ in whose entanglement we are
interested. The first CM must fulfill
 \be
 \label{symmetry}
 \Gamma=\frac1{|G|}\sum_{g\in G}(T_g\oplus T_g)\Gamma
 (T_g\oplus T_g)^T.
 \ee

The CM $\gamma$ of a two-mode subsystem can always be written, up
to local symplectic transformations $S_{1,2}$, in the standard
form \cite{normform}
 \begin{equation}\label{SimonNormalform}
 (S_1\oplus S_2)\gamma(S_1\oplus S_2)^T=\begin{pmatrix}
  n_A & k_q \\
  k_q & n_B
 \end{pmatrix}\oplus\begin{pmatrix}
  n_A & k_p \\
  k_p & n_B
 \end{pmatrix}.
 \end{equation}
The fact that $G$ contains by assumption an element which maps
$k\leftrightarrow l$ immediately implies that $n_A=n_B$ and
$S_1=S_2=S$. Hence, given a global CM we can always find another
one given by $\big(\bigoplus_{i=1}^N
S\big)\Gamma\big(\bigoplus_{i=1}^N S\big)^T$, which is also
$G$--symmetric, and such that $\gamma$ has the standard form
(\ref{SimonNormalform}) with $n_A=n_B=:n$.

We are interested in maximizing the EoF of $\gamma$. Since
$n_A=n_B$ we can use the results of \cite{GWKWC03}, which show
that this quantity is given by $E_F(\Delta)$, where the function
$E_F$ has been given in (\ref{EoF}), which is a monotonically
decreasing function of the so--called {\it EPR--uncertainty}
$\Delta$. Thus, maximizing the EoF is equivalent to minimizing
$\Delta$. This last quantity is a highly nonlinear function of the
parameters $n,k_q,k_p$, and thus minimizing it with respect to all
possible global $\Gamma$ looks as a very daunting task. In order
to overcome this problem, the trick is to linearize the expression
of $\Delta$ by including an extra maximization in the problem,
i.e. writing
 \bma
 \bea
 \label{EPR1}
 \Delta &=& \inf_{s>0} \tr{\gamma \Big(s h_+^{(k,l)}\oplus \frac1s
 h_-^{(k,l)}\Big)}\\
 &=& \inf_{s>0} \tr{\Gamma \Big(s H_+\oplus \frac1s H_-\Big)}.
 \eea
 \ema
In the last step we have used that $\gamma$ is the reduced CM of
$\Gamma$, and Eq.(\ref{symmetry}).

We show now that $\Gamma$ must correspond to a pure state. If a CM
$\Gamma_m$ corresponds to a $G$-symmetric mixed state, it can
always be decomposed into a $G$-symmetric pure state CM $\Gamma_p$
and a $G$-symmetric matrix $M\geq 0$ via $\Gamma_m=\Gamma_p + M$
\footnote{To see this, note that we can always subtract a  matrix
$\tilde{M}\geq 0$ from $\Gamma_m$ such that still $(\Gamma_m -
\tilde{M})\geq i\sigma$ holds. Averaging this equation over the
group $G$ tells us that we can always choose $\tilde{M}$ to be
invariant under $G$. Subtracting as much of a $G$-symmetric and
positive semi-definite matrix as possible leads us then to the
symmetric pure state CM $\Gamma_p$.}. This decomposition can be
interpreted as adding classical Gaussian noise to $\Gamma_p$
\cite{WW01a}. Since this will certainly not increase the
entanglement between any two modes of the system,  maximal
entanglement will be attained if the overall CM $\Gamma$
corresponds to a pure state.

We can thus exploit the fact that every pure state CM can be
written as
 \begin{equation}\label{GXY}
 \Gamma=\left(\begin{array}{cc}
  X & XY \\
  YX\  & YXY+X^{-1}
 \end{array}\right)\;,
 \end{equation}
with $X>0$ and $Y=Y^T$ \cite{WGKWC03}. Hence, the EPR-uncertainty
(\ref{EPR1}) becomes
 \begin{equation}
 \label{EPR2}
 \Delta=\inf_{s>0}s\tr{X H_+}+\frac1s \tr{(X^{-1}+YXY) H_-},
 \end{equation}
Maximizing the entanglement means  minimizing $\Delta$ with
respect to $X$ and $Y$ under the constraint that they parameterize
a $G$-symmetric CM. We can, however, drop this constraint since
the symmetry of $H_{\qp}$ will force the optimal $\Gamma_0$ to
have the right symmetry. Moreover, $\Gamma_0$ will be the ground
state corresponding to the Hamiltonian matrix $H=H_+\oplus H_-$
since
 \begin{eqnarray}
 \label{final1}
 \Delta_0&=&\inf_{X,Y}\Delta=\inf_{X} \tr{X
 H_+}+\tr{X^{-1} H_-}\\
 \label{final2}
 &=&\inf_{\Gamma}\tr{\Gamma(H_+\oplus H_-)}
 \end{eqnarray}
where we have first set $Y=0$, since $\tr{YXYH_-}\geq 0$ and then
incorporated the infimum over $s$ into that over $X$. This
completes the proof of our main result, since $\Delta_0$ is by
Eqs.(\ref{Hqptwirl},\ref{final2}) equal to the ground state energy
$E_0$ of the Hamiltonian in Eq.(\ref{Hamiltonian}).

By imposing less restrictions on the symmetry group than requiring
the existence of an element which interchanges $k\leftrightarrow
l$, the maximal achievable EoF could grow. In the examples on
chains and lattices, we implicitly imposed the translational
\emph{and} reflection symmetry. The same results can however be
derived with only the translational symmetry. In fact, the
presented proof can be extended in a straight forward manner to
all symmetry groups with Abelian commutant, including those of
rings and cubic lattices, without imposing reflection symmetry.
The proof can be found in the appendix.

 \vspace*{0pt}In
conclusion, we have determined the maximal entanglement between
two modes under the constraint that the overall system is in a
Gaussian state which has some symmetry with respect to the
ordering of the modes. The result was derived by linearizing the
entanglement functional which permits to perform the maximization
in a simple way. We find that the maximal entanglement is
connected to the ground state of a particular quadratic
Hamiltonian which possesses the same symmetry as the state. The
state that maximizes the EoF is precisely the ground state of such
a Hamiltonian. The maximal entanglement turned out to be finite in
all the discussed cases, and is even comparable to the
(conjectured) values for spin $\frac12$ systems for the case of
rings \cite{OW01}. We have shown how the entanglement decreases
with the number of spatial dimensions, and how it depends on the
geometry of the state. Finally, although we have concentrated here
on Gaussian states, we think that similar results may be obtained
for the case of qubits. In that case it is also possible to
linearize the expression for the EoF \cite{inpreparation} and in
this way to relate it to the ground state energy of a specific
nearest--neighbor interaction Hamiltonian, which allowed us to
solve the problem for Gaussian states.

\vspace*{-1pt}\acknowledgements We acknowledge interesting
discussions with J. Garcia Ripoll and G.Giedke. This work was
supported in part by the E.C. (projects RESQ and QUPRODIS) and the
Kompetenznetzwerk ``Quanteninformationsverarbeitung'' der
Bayerischen Staatsregierung.


\appendix\section{Appendix: groups with Abelian commutant}
In this appendix, it will be shown that the reflection symmetry
need not be imposed if the commutant of the group is Abelian, as
the states with maximal EoF automatically have this property. The
proof is as follows. The reduced covariance matrix corresponding
to modes $k$ and $l$ of a state with a symmetry that maps $g(k)=l$
is of the form\vspace*{-5pt}
\[\gamma=\begin{pmatrix}
  a & b & d & e \\
  b & c & f & g \\
  d & f & a & b \\
  e & g & b & c \\
\end{pmatrix}.\]
It can readily be checked that this can be brought into symmetric
normal form $\tilde{\gamma}$ by local symplectic transformations
of the form
\[\tilde{\gamma}=(O\oplus \id)(S\oplus S)\gamma(S\oplus S)^T(O^T\oplus \id)\]
with $S$ symplectic and $O$ orthogonal. Suppose now that $\gamma$
and the global covariance matrix $\Gamma$ are such that $S=\id$,
which can always be done by a transformation $\Gamma\mapsto
(\oplus_{i=1}^N S')\Gamma(\oplus_{i=1}^N S')^T$. Then the
EPR-uncertainty $\Delta$ is given by\vspace*{-5pt}
\[\Delta=\inf_{s>0}{\rm Tr}\left[\gamma \underbrace{(O^T\oplus \id)\left(%
\begin{array}{cccc}
  s & 0 & s & 0 \\
  0 & 1/s & 0 & -1/s \\
  s & 0 & s & 0 \\
  0 & -1/s & 0 & 1/s \\
\end{array}%
\right)(O\oplus \id)}_{h}\right].\] Parameterizing $O=\left(%
\begin{array}{cc}
  \cos(\theta) & -\sin(\theta) \\
  \sin(\theta) & \cos(\theta) \\
\end{array}%
\right),
$ one gets
\begin{eqnarray} h_{qq}&=&\left(%
\begin{array}{cc}
  s\cos^2(\theta)+\frac{\sin^2(\theta)}{s} & s\cos(\theta) \\
  s\cos(\theta) & s \\
\end{array}%
\right)\\
h_{qp}=h_{pq}^T&=&\left(%
\begin{array}{cc}
  \left(\frac{1}{s}-s\right)\cos(\theta)\sin(\theta) & -\frac{\sin(\theta)}{s} \\
  -s\sin(\theta) & 0 \\
\end{array}%
\right)\\
h_{pp}&=&\left(%
\begin{array}{cc}
  s\sin^2(\theta)+\frac{\cos^2(\theta)}{s} & -\frac{\cos(\theta)}{s} \\
  -\frac{\cos(\theta)}{s} & \frac{1}{s} \\
\end{array}%
\right)
\end{eqnarray}
Let's next extend these Hamiltonian blocks to $N$-modes by
averaging over the group:
\[H_{\alpha\beta}=\frac1{|G|}\sum_{g\in G} T_gh_{\alpha\beta}T_g^T\]
The goal is now to calculate the sum of the symplectic eigenvalues
of this Hamiltonian. As the commutant of the group was assumed to
be Abelian, all $N\times N$ matrices
\[A_{ij}=\frac1{|G|}\sum_{g\in G}
T_g|i\rangle\langle j|T_g^T,\] $(i,j)\in \{0,1\}$, commute and
hence can be diagonalized simultaneously by a unitary operator
(eventually complex), yielding diagonal elements
$\lambda_{ij}^\mu$, where $(i,j)\in \{0,1\}$ and $1\leq\mu\leq N$.
Applying the same unitary transformation to the four matrices
$H_{\alpha\beta}$, the complete Hamiltonian becomes a direct sum
of $2\times 2$ blocks of the form
\begin{equation}\label{isdjf}B_\mu=\left(%
\begin{array}{cc}
  b_{qq}^{\mu} & b_{qp}^{\mu} \\
  {b_{qp}^{\mu}}^* & b_{pp}^{\mu}
\end{array}%
\right)\;,\quad b_{\alpha\beta}^{\mu}=\sum_{ij}
h_{\alpha\beta}^{ij}\lambda_{ij}^{\mu}\;.\end{equation}

In general, the symplectic eigenvalues of a positive operator
$\Gamma$ are given by the square roots of the eigenvalues of the
operator $(\Gamma \sigma\Gamma^\dagger \sigma^T)$, and the unitary
operation under consideration does not change these eigenvalues.
It can now readily be checked that the sum of the symplectic
eigenvalues corresponding to the block $B_\mu$ is given by the
formula
\[\nu^\mu_1+\nu_2^\mu=\sqrt{b_{qq}^{\mu} b_{pp}^{\mu}-({\rm Re}(b_{qp}^{\mu}))^2}.\]
Due to equation (\ref{isdjf}), it holds that
\[b_{qq}^{\mu} b_{pp}^{\mu}-({\rm Re}(b_{qp}^{\mu}))^2=(\bar{\lambda}^\mu)^T
Q\bar{\lambda}^\mu\] where
\begin{eqnarray*}
\bar{\lambda}^{\mu}&=&\left(%
\begin{array}{c}
  \lambda_{00}^\mu \\
  {\rm Re}(\lambda_{01}^\mu) \\
  \lambda_{11}^\mu \\
\end{array}%
\right),\\
Q&=&\left(%
\begin{array}{ccc}
  1 & 0 & 1 \\
  0 & -4 & 0 \\
  1 & 0 & 1 \\
\end{array}%
\right)+\frac{\sin^2(\theta)}{2}\left(s-\frac{1}{s}\right)^2\left(%
\begin{array}{ccc}
  0 & 0 & 1 \\
  0 & -2 & 0 \\
  1 & 0 & 0 \\
\end{array}%
\right).
\end{eqnarray*}
The presence of the second term in $Q$ (which is not there when
reflection symmetry is present) can never lead to an improvement
of the entanglement, as it can only increase the symplectic
eigenvalues. Indeed, it holds that
\[\lambda_{00}^\mu\lambda_{11}^\mu-|\lambda_{01}^\mu|^2\geq 0,\]
as it is the determinant of a principal submatrix of a positive
$2N\times 2N$ matrix $P$. ($P$ is obtained by first applying the
group symmetry to the hypothetical blocks $h_{qq}=|0\rangle\langle
0|;h_{qp}=h_{pq}^T=|0\rangle\langle 1|;h_{pp}=|1\rangle\langle 1|$
(which is manifestly positive), and then diagonalizing the four
blocks by a unitary transformation.) Therefore, the optimal choice
for the orthogonal matrix $O$ is to choose it equal to the
identity, imposing that the optimal solution obeys the reflection
symmetry.

\end{document}